\documentclass[twocolumn,english,prl]{revtex4-1}
\usepackage[T1]{fontenc}
\usepackage[latin9]{inputenc}
\usepackage{float}
\usepackage{amsmath}
\usepackage{graphicx}

\makeatletter
\usepackage{babel}
\usepackage{babel}

\usepackage{babel}

\usepackage{babel}

\makeatother

\usepackage{babel}
\begin{document}

\title{Two diverging length scales in the structure of jammed packings}

\author{Daniel Hexner}
\email{danielhe2@uchicago.edu}

\selectlanguage{english}%

\affiliation{The James Franck Institute and Department of Physics, The University
of Chicago, Chicago, IL 60637, USA and Department of Physics and Astronomy,
The University of Pennsylvania, Philadelphia, PA, 19104, USA}

\author{Andrea J. Liu}

\affiliation{Department of Physics and Astronomy, The University of Pennsylvania,
Philadelphia, PA, 19104, USA}

\author{Sidney R. Nagel}

\affiliation{The James Franck and Enrico Fermi Institutes and The Department of
Physics, The University of Chicago, Chicago, IL 60637, USA}
\begin{abstract}
At densities higher than the jamming transition for athermal, frictionless
repulsive spheres we find two distinct length scales, both of which
diverge as a power law as the transition is approached. The first,
$\xi_{Z}$, is associated with the two-point correlation function
for the number of contacts on two particles as a function of the particle
separation. The second, $\xi_{f}$, is associated with contact-number
fluctuations in subsystems of different sizes. On scales below $\xi_{f}$
the fluctuations are highly suppressed, similar to the phenomenon
of hyperuniformity usually associated with density fluctuations. The
exponents for the divergence of $\xi_{Z}$ and $\xi_{f}$ are different
and appear to be different in two and three dimensions. 
\end{abstract}
\maketitle
A key signature of a critical phase transition is the existence of
a correlation length, $\xi$, which diverges at the critical point.
On scales smaller than $\xi$ the constituents act in a cooperative
manner, while on large scales the system typically behaves as if it
were non-interacting \cite{Wilson1975,kadanoff2000statistical}. The
correlation length is defined by the second moment of the two-point
correlation function for the local order parameter. For nonequilibrium
disordered transitions, however, the local order parameter is not
always known.

The jamming transition of a system of soft repulsive spheres is an
example of such a transition. It occurs at temperature $T=0$ as the
applied pressure (or packing fraction) is increased driving the system
from a floppy to a rigid state. While various length scales have been
shown to diverge as the jamming critical point is approached, they
do not characterize the structure itself but rather the normal modes,
the mechanical stability, and the elastic response of the system \cite{Silbert2005,Wyart2005_epl,dipole_response,Goodrich2013,Schoenholz2013}.

In this paper, we show that the onset of rigidity is associated with
the divergence of two distinct structural length scales, $\xi_{Z}$
and $\xi_{f}$, both associated with the \emph{contact number}. The
contact number, $Z_{i}$, is the number of neighbors with which a
particle $i$ interacts and varies from one particle to the next.
One of these lengths, $\xi_{Z}$, is associated with the decay of
the two-point spatial correlation function for $Z$. Our finding that
$\xi_{Z}$ diverges at the jamming transition motivates us to examine
the size of contact number fluctuations in subsystems of different
sizes. In contrast to usual behavior of critical points, where the
long-range correlations result in diverging fluctuations, here we
find that the contact number fluctuations are highly suppressed on
\emph{large scales}. Namely, at the jamming transition the contact
fluctuations in a volume of $\ell^{d}$ scale as its surface $\ell^{d-1}$,
which is the smallest possible scaling consistent with local randomness
in the contact network. Thus, a system at the jamming transition appears
to have \emph{contact hyperuniformity}, a term we introduce in analogy
to the concept of density hyperuniformity~\cite{torquato_local_2003},
which was first observed in the mass distribution in the early universe
and in plasmas \cite{Jancovici1981_onecomponent_plasma,Levesque2000_onecomponent_plasma,Gabrielli2002_cosmolo}.
We note that similar generalizations of hyperuniformity have been
introduced, for example, in the study of foams~\cite{Roth2013},
pattern formation\cite{Torquato2018} and random fields\cite{Torquato2018}.
At a finite distance from the jamming transition the hyperuniform
scaling persists up to a finite distance $\ell<\xi_{f}$, where $\xi_{f}$
diverges at the jamming transition. We show how the exponents characterizing
the divergence at the jamming transition of $\xi_{f}$ and of $\xi_{Z}$
are related; surprisingly, both exponents appear to depend on the
dimension of the system, $d$ in contrast to previously observed lengths
that diverge with dimension-independent behavior \cite{Silbert2005,Wyart2005_epl,dipole_response}.

Our analysis is based on numerically generated packings in either
$d=2$ spatial dimensions with $N=$128,000 polydisperse, or $d=3$
with $N=10^{6}$ monodisperse frictionless soft repulsive particles
in a volume, $V$. The harmonic repulsion between particles is given
by: 
\begin{equation}
U(r_{ij})=\frac{1}{2}\epsilon\left(1-\frac{r_{ij}}{\sigma_{ij}}\right)^{2}\Theta\left(1-\frac{r_{ij}}{\sigma_{ij}}\right),\label{eq:hertzian}
\end{equation}
where $\epsilon$ is the characteristic energy, $\Theta(x)$ is the
Heaviside step function, and $r_{ij}$ and $\sigma_{ij}$ are respectively
the separation between particles $i$ and $j$ and sum of their radii.
Configurations are prepared by standard methods used for studies of
jamming \cite{Review,Ohern}; spheres are distributed randomly in
space and the system's energy is minimized using the FIRE\cite{FIRE}
algorithm to produce a zero-temperature jammed configuration where
force balance is maintained on every particle~\cite{Goodrichfinitesize}.

The simplest measure of correlations is a two-point correlation function.
To this end, we define $\delta Z_{i}\equiv Z_{i}-\overline{Z}$, measuring
the deviation of $Z_{i}$ from its average $\overline{Z}=\frac{1}{N}\sum_{i}Z_{i}$
and its two-point correlation function $h_{Z}\left(r\right)=\left\langle \delta Z\left(r\right)\delta Z\left(0\right)\right\rangle $.
Here, $\delta Z\left(r\right)=\sum_{i}\delta Z_{i}\delta\left(r-r_{i}\right)$
where $r_{i}$ denotes the location of the particles center and the
average is over different realizations and all equidistant locations
in the packing. For a finite number of realizations, $N_{0}$, $h_{Z}\left(r\right)$
can be determined to an accuracy of $1/\sqrt{N_{0}}$ . Since $h_{Z}\left(r\right)$
decays to zero as a function of distance, a growing number of realizations
are needed to measure it when $r$ is large. We therefore measure
the contact-number structure factor

\begin{equation}
S_{Z}\left(q\right)=\frac{1}{N}\left\langle \left|\sum_{i=1}^{N}\delta Z_{i}e^{-iqr_{i}}\right|^{2}\right\rangle .\label{eq:dZq}
\end{equation}
Here the average is over different realizations and directions of
the wave vector $q$. The relation between $S_{Z}\left(q\right)$
and $h_{Z}\left(r\right)$ can be made apparent by using $\rho,$
the particle density, and the definition of $\delta Z\left(r\right),$
which yields: 
\begin{align}
S_{Z}\left(q\right)= & \left\langle \left(\delta Z_{i}\right)^{2}\right\rangle \label{eq:corr}\\
 & +\frac{1}{\rho}\int d^{d}r\,h_{Z}\left(r,0\right)e^{-iqr}.\nonumber 
\end{align}

At the jamming transition in the thermodynamic limit, $\overline{Z}=Z_{c}=2d$
(Here we do not include rattlers where $Z_{i}$ is too small to confine
a particle rigidly.). This corresponds to the minimal number of contacts
needed for rigidity and $\Delta Z\equiv\overline{Z}-Z_{c}$ measures
the distance from the critical point \cite{Durian1995,Ohern,Wyart2005Ann}.
Correlations in $\Delta Z$ have been previously suggested \cite{Wyart2005_compression,Moukarzel2012}
but their nature was not studied.

\begin{figure}[H]
\includegraphics[scale=0.6]{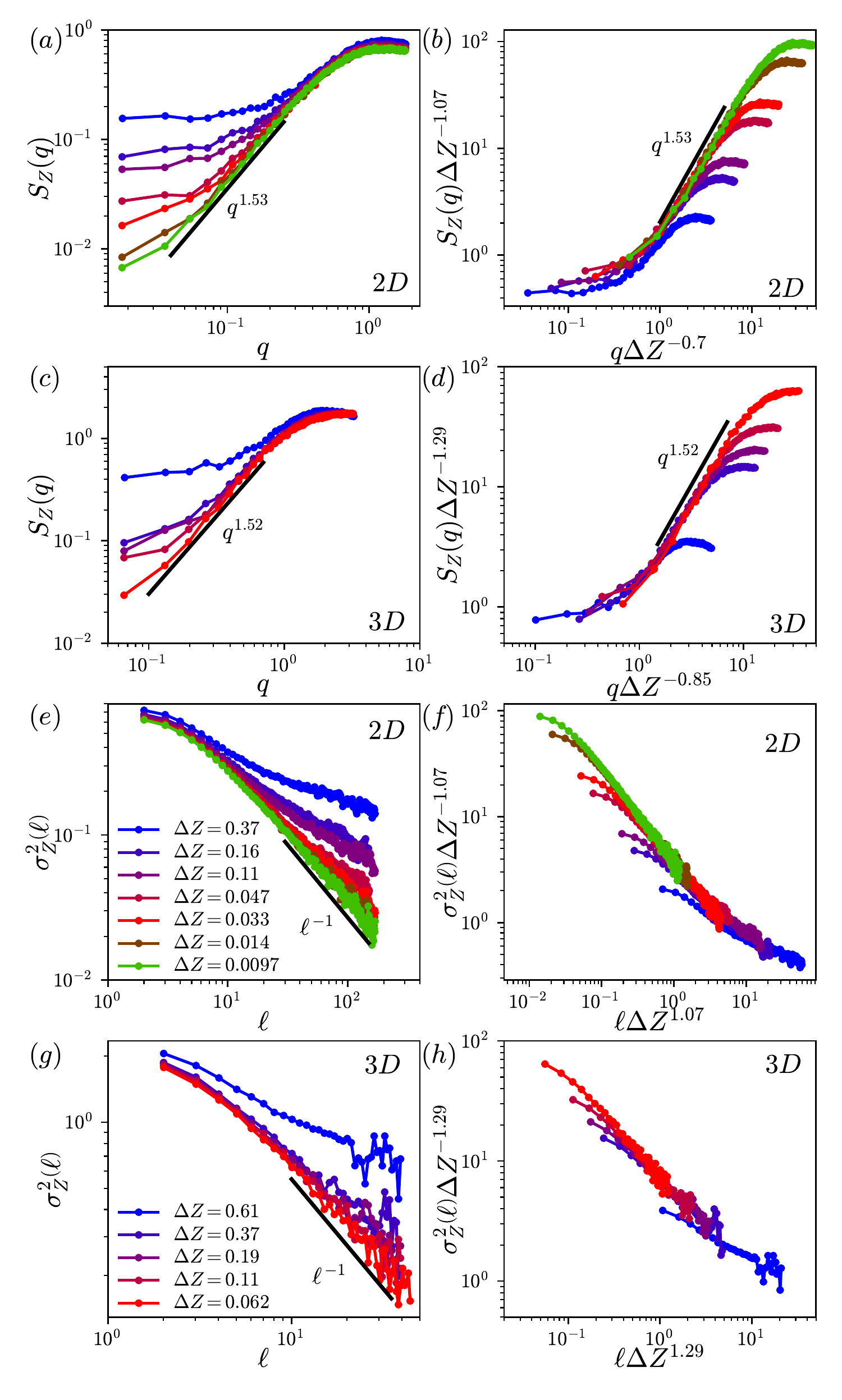}

\caption{The length-dependent fluctuation of the contact number at different
values of $\Delta Z$. The real-space fluctuations are shown in (a)
$d=$2 and (c) $d=$3 and the fluctuations in Fourier space in (e)
$d=$2 and (g) $d=$3. The values of $\Delta Z$ are shown in the
legend. (b),(d),(f) and (h) show the collapse of the data. The number
of particles in 2d is $N=128,000$ and in 3d is $N=10^{6}$ . \label{fig: dz_fluc}}
\end{figure}

Fig.~\ref{fig: dz_fluc}a and c show $S_{Z}\left(q\right)$ in $d=2$
and $d=3$, for different values of $\Delta Z$. As $q\rightarrow0$,
$S_{Z}\left(q\right)$ approaches a constant that depends on $\Delta Z$.
At intermediate values of $q$, this function rises steeply, approximately
as a power-law, $S_{Z}\left(q\right)\propto q^{\alpha}$, where $\alpha_{2d}=1.53\pm0.04$
and $\alpha_{3d}=1.52\pm0.05$. The regime of $q\gtrsim1$, corresponding
to wavevectors greater than the inverse particle diameter, is not
the focus of this paper. In the limit of $\Delta Z\rightarrow0$,
the power-law regime $q^{\alpha}$, appears to extend to arbitrarily
small q-values, implying that $S_{Z}\left(q\rightarrow0\right)=0$.
Below we will argue that this has important consequences for large-scale
contact fluctuations. In this limit, the real-space correlation function
decays as a power-law with an exponent that is fairly large: $h_{Z}\left(r\right)\propto-r^{-d-\alpha}$.
The negative sign can be inferred from the fact that the first term
in Eq. \ref{eq:corr} is positive and can only be reduced if $h_{Z}\left(r\right)<0$
at large distances.

The transition between the first two regimes defines a length scale,
$\xi_{Z}=2\pi/q_{c}$ where $q_{c}$ is the crossover wave-vector.
This length scale diverges as $\Delta Z\rightarrow0$, presumably
in a power-law manner $\xi_{Z}=\Delta Z^{-\nu_{Z}}$. To measure this
length scale it is convenient to write $S_{Z}\left(q\right)$ in the
form of a scaling function: 
\begin{equation}
S_{Z}\left(q\right)=\Delta Z^{\beta}f\left(q\xi_{Z}\right).
\end{equation}
This implies that the data can be collapsed by rescaling the x-axis
by $\Delta Z^{-\nu_{Z}}$ and the y-axis by $\Delta Z^{-\beta}$.
To further constrain $\nu_{Z}$ and $\beta$ we note that $f\left(x\right)$
has two limiting behaviors: 
\begin{equation}
f\left(x\right)=\begin{cases}
const & x\ll1\\
x^{\alpha} & x\gg1
\end{cases}.
\end{equation}
This scaling regime is cut off when $q^{-1}$ becomes of the order
of several particle diameters. In the limit of $q\xi_{Z}\gg1$, $S_{Z}\left(q\right)$
is independent of $\Delta Z$ implying that 
\begin{equation}
\beta=\alpha\nu_{Z}.\label{eq:scaling}
\end{equation}

Thus by measuring $\alpha$ the data can be collapsed by varying a
single exponent. Fig. \ref{fig: dz_fluc}b and d shows the collapse
for both two and three dimensions, where the best collapse is found
for $\nu_{Z}^{2d}=0.7_{-0.1}^{+0.05}$ and $\nu_{Z}^{3d}=0.85_{-0.1}^{+0.15}$.
The errors arise from the uncertainty in $\alpha$ and the finite
range of the data. Our results suggest that $\nu_{Z}$ may be different
in two and three dimensions, in contrast to other critical exponents
associated with jamming which do not appear to depend on dimension.
We note that we cannot rule out that this apparent difference arises
due to corrections to scaling near the upper-critical dimension, thought
to be two dimensions in this case\cite{WyartReview}. 

We turn next to consider what $S_{Z}\left(q\right)$ implies for the
large-scale behavior of the contact fluctuations. We first note that
previously-studied \emph{density} hyperuniformity can be measured
from the low-$q$ behavior of the density structure factor, $S_{\rho}\left(q\right)=\frac{1}{N}\left|\sum_{i}e^{-iqr_{i}}\right|^{2}$.
The low-$q$ limit describes long length-scale density fluctuations\cite{hansen2006theory}:
$S_{\rho}\left(q\rightarrow0\right)=\left(\left\langle N^{2}\right\rangle -\left\langle N\right\rangle ^{2}\right)/\left\langle N\right\rangle $,
where $N$ is the number of particles in the system. Therefore, if
$S_{\rho}\left(q\rightarrow0\right)=0$ the density fluctuations are
sub-extensive and suppressed compared to typical equilibrium systems~\cite{torquato_local_2003},
on par with those of a perfect crystal. Our result that $S_{Z}\left(q\right)\rightarrow q^{\alpha}$
at low $q$ at the jamming transition implies that at the transition,
the system obeys \emph{c}ontact hyperuniformity: the contact fluctuations
are highly suppressed at long length scales despite the local randomness
in $Z_{i}$.

To take a closer look at contact hyperuniformity, we measure contact
fluctuations as a function of length scale. We consider a sub-region
with linear dimension $\ell$, specifically a hypercube of volume
$\ell^{d}$ in $d$ dimensions. The fluctuations of $\delta Z_{i}=Z_{i}-\overline{Z}$
in hypercubes of this size are characterized by: 
\begin{equation}
\sigma_{Z}^{2}\left(\ell\right)=\frac{1}{\ell^{d}}\left\langle \left(\sum_{i\in\ell^{d}}\delta Z_{i}\right)^{2}\right\rangle .\label{eq:dZreal_space}
\end{equation}
where the angular brackets denote an average over different sub-regions
(of size $\ell$) in a given packing as well as different realizations.

If the $\delta Z_{i}$ were uncorrelated random variables then $\sigma_{Z}^{2}\left(\ell\right)\propto const$
and any deviations from this would imply correlations. Figure~\ref{fig: dz_fluc}e
shows $\sigma_{Z}^{2}\left(\ell\right)$ in $d=2$ for different values
of $\Delta Z$. Figure~\ref{fig: dz_fluc}g shows the results are
qualitatively similar in $d=3$. At the smallest value of $\Delta Z$,
$\sigma_{Z}^{2}$ approaches $\ell^{-1}$ at large $\ell$. This implies
that fluctuations in the contact number are suppressed; there must
be correlations in $Z_{i}$ to insure this property as seen in $S_{Z}\left(q\right)$.
Surface fluctuations are inevitable since translating slightly the
measurement window varies which particles are within the measurement
window, and as result also the contact number. Increasing $\Delta Z$
shows that at large $\ell$ there is a crossover from $\ell^{-1}$
to what we will argue is a constant. This implies Poissonian fluctuations
in this regime. The cross-over between these two behaviors defines
a length scale, $\xi_{f}$, which diverges as $\Delta Z\rightarrow0$.
Strangely, this crossover appears to be very slow in comparison to
the data presented for $S_{Z}\left(q\right)$. We will argue that
$\xi_{f}$ is indeed larger than $\xi_{Z}$ , diverging faster than
the latter.

We now argue that $\xi_{f}$ diverges with an exponent that is different
from $\xi_{Z}$. To relate $\xi_{f}$ to the measured exponents in
$S_{Z}\left(q\right)$ it is useful to express $\sigma_{Z}^{2}\left(\ell\right)$
in terms of the two-point correlation function. Using, $h_{Z}\left(r_{2},r_{1}\right)=\left\langle \delta Z\left(r_{2}\right)\delta Z\left(r_{1}\right)\right\rangle $
it straightforward to show that:

\begin{equation}
\sigma_{Z}^{2}\left(\ell\right)=\rho\left\langle \left(\delta Z_{i}\right)^{2}\right\rangle +\frac{1}{\ell^{d}}\int_{\ell^{d}}d^{d}r_{1}\int_{\ell^{d}}d^{d}r_{2}h_{Z}\left(r_{2},r_{1}\right),\label{eq:sigma_h_z}
\end{equation}
where $\rho$ is the density of particles and the integral is over
the hyper-cube. In the limit of $\ell\rightarrow\infty$, surface
terms can be neglected which leads to $\sigma_{Z}^{2}\left(\ell\rightarrow\infty\right)=\rho\left\langle \left(\delta Z_{i}\right)^{2}\right\rangle +\int d^{d}r\,h_{Z}\left(r,0\right)$.
This is also equal to $S_{Z}\left(q\rightarrow0\right)/\rho$ (see
Eq. \ref{eq:corr}) such that if $S_{Z}\left(q\right)\propto q^{\alpha}$
on all length scales then fluctuations are sub-extensive, $\sigma_{Z}^{2}\left(\ell\rightarrow\infty\right)=0$.

On finite scales the relation is more subtle, leading to two distinct
length scales. Ref. \cite{Torq_derivation_q} considers the relation
between the structure factor and scaling of the density fluctuations
as a function of scale. We apply their analysis here to the contact
statistic, and find that if $S_{Z}\left(q\right)\propto q^{\alpha}$
then asymptotically $\sigma_{Z}^{2}\left(\ell\right)\propto\ell^{-\psi}$
where, 
\begin{equation}
\psi=\begin{cases}
\alpha & \alpha<1\\
1 & \alpha>1
\end{cases}.
\end{equation}
This non-analytic relation arises because the fluctuations cannot
decay faster that $\ell^{-1}$ \textendash{} the contribution due
to fluctuations on the surface. Since in our case $\alpha>1$ we expect
that near the jamming transition $\sigma_{Z}^{2}\left(\ell\right)\propto\ell^{-1}$
in agreement with the data in Figs.~\ref{fig: dz_fluc}e and g. The
exponent $\nu_{f}$ can be estimated by comparing $\ell^{-1}$ to
the asymptotic behavior $\sigma_{Z}^{2}\left(\ell\rightarrow\infty\right)=S_{Z}\left(q\rightarrow0\right)/\rho\propto\Delta Z^{\beta}$,
yielding $\nu_{f}=\beta$. Using the values of $\beta$ obtained in
the collapse of $S_{Z}\left(q\right)$ we find that $\nu_{f}^{2d}=1.07_{-0.18}^{+0.1}$
and $\nu_{f}^{3d}=1.29_{-0.19}^{+0.27}$. Figure~\ref{fig: dz_fluc}f
and h show that these exponents provide a reasonable collapse of $\sigma_{Z}^{2}\left(\ell\right)$.
Thus, as we asserted above, the fluctuation length scale diverges
with an exponent different from that of the correlation length. We
argue that generically for systems that have suppressed fluctuations
there are two distinct length scales satisfying $\nu_{f}>\nu_{Z}$,
when $\alpha>1$, and a single length scale $\nu_{f}=\nu_{Z}$ when
$\alpha<1$.

In summary, we have shown that there are two diverging length scales
which characterize the contact fluctuations near the jamming transition.
Unlike traditional equilibrium critical phenomena, the diverging length
scale in the two-point correlations of $\Delta Z$ is not accompanied
with large fluctuations but rather with the suppression of contact
fluctuations on large scales. Indeed it is precisely this \emph{smallness}
of fluctuations that make this structural ``order'' elusive, as
there are no large scale features seen to the naked eye.

The small fluctuations in $Z_{i}$ suggest that it should be considered
a control parameter, analogous to temperature in the Ising model,
rather than as an order parameter. If we adopt this view, the Harris
criterion compares the average of the control parameter, $\Delta Z$
inside a volume $\xi_{f}^{d}$ to its fluctuations. Stability requires
that the average must vanish faster than the fluctuations. The average
coordination number scales as $\Delta Z$ while contact hyperuniformity
implies that the fluctuations scale as the surface area of the region
of size $\xi_{f}$, as our simulations suggest. The magnitude of the
fluctuations scale as the square root of the variance, namely $\xi_{f}^{-\left(d+1\right)/2}$.
Comparing these, we obtain the inequality 
\begin{equation}
\nu_{f}>\frac{2}{d+1}.\label{eq:in_equality}
\end{equation}
The fact that our observed values obey this inequality in $d=2$ and
$d=3$ suggests that $\Delta Z$ should indeed be viewed as the control
variable rather than an order parameter. We note that while the Harris
criterion is usually employed in disordered systems in which fluctuations
in the control parameter are quenched, here the fluctuations $Z_{i}$
emerge from many-body interactions.

This conclusion is consistent with the choice made in the scaling
ansatz for the jamming transition~\cite{Goodrich2016}, which suggests
that packing fraction and shear strain should be considered the order
parameters. However, there are no apparent diverging length scales
in the two-point correlations of the packing fraction~\cite{Ikeda2017}.
A single contact connects two particles and is therefore related to
the two-point density correlation function. Therefore, the two-point
contact correlations studied here correspond to four-point density
correlations. Our results demonstrate that while ``order'' can sometimes
be found in plain sight, its identification, especially in disordered
systems, may require a carefully tailored higher-order correlation
function, as has been proposed for glasses~\cite{Bouchaud2005,Albert2016}.

Our results on contact hyperuniformity should be compared to recent
studies of density fluctuations at or above the jamming transition.
It has been suggested that systems at and above the jamming transition
are hyperuniform in density\cite{Donev_2005,Zachary2011,Silbert2009,Weeks2010,Weeks2011,Berthier2011},
but this is controversial. Recent studies suggest that systems are
not hyperuniform in density above above the jamming transition\cite{Ikeda2017,Teitel2015,Berthier2015,Torquato2016};
states prepared upon approach to jamming from below appear to be even
less hyperuniform with better equilibration \cite{Ozawa2017}. However,
studies of very large systems that explicitly identify a crossover
length $\xi_{\rho}$ below which the system is uniform\cite{Durian2017}
suggest that $\xi_{\rho}$ might increase somewhat in the dual limit
as the pressure is decreased towards the jamming transition and the
equilibration time increases\cite{Durian2017}. Density hyperuniformity
has also been predicted for sedimentation\cite{Goldfriend2017} and
periodically sheared suspensions\cite{HexnerHyper2015,tjhung2015hyperuniform}.



Our findings open the door to studying several aspects of the jamming
transition. \emph{Dynamics:} In studying jamming dynamics, our spatial
metrics could be used study how spatial ``order'' evolves as the
spheres approach the jammed state. This is characterized by a dynamical
exponent relating relaxation time to the correlation length, $\tau\propto\left(\xi_{Z}\right)^{\mu}$.
Such an exponent was identified in the first study of the jamming
transition by Durian~\cite{Durian1995}. \emph{Interplay of structure
and elasticity:}We expect that $\xi_{Z}$ and $\xi_{f}$ should be
reflected in the diverging length scales found in elasticity~\cite{Silbert2005,Maloney2015,Tighe_periodic_forcing,dipole_response}.
We note that Ref. \cite{Tighe_periodic_forcing} finds a length scale
that diverges as $\Delta Z^{-0.66}$ in two dimensions, consistent
with the length scale found in $S_{Z}\left(q\right)$\footnote{They also suggest that a linear combination of $\Delta Z^{-0.5}$
and $\Delta Z^{-1.0}$ fits as well if they weighted appropriately
. Ref. \cite{Maloney2015} measures the same length scale and finds
that it diverges as $\Delta Z^{-0.8}$. We have repeated their analysis
and our data is more consistent with $\Delta Z^{-0.66}$. }. \emph{Role of dimensionality:} Our results suggest that some of
the exponents depend on dimension, in contrast to previous findings.
This suggests that contact hyperuniformity has a non-mean-field flavor
in low dimension. It would be interesting to examine this length scale
in mean-field calculations. 
\begin{acknowledgments}
We thank D. Durian, J. Kurchan, D. Levine, T. A. Witten, N. Xu and
F. Zamponi for instructive discussions. We gratefully acknowledge
support from the Simons Foundation for the collaboration ``Cracking
the Glass Problem\textquotedblright{} award \#348125 (DH, SRN) and
\#327939 (AJL), the US Department of Energy, Office of Basic Energy
Sciences, Division of Materials Sciences and Engineering under Award
DE-FG02- 05ER46199 (AJL) and DE-FG02-03ER46088 (SRN), and the University
of Chicago MRSEC NSF DMR-1420709 (DH). We gratefully acknowledge the
computer resources enabled by the RCC Legacy Simulation Allocation
Program. 
\end{acknowledgments}

\bibliographystyle{apsrev4-1}
\bibliography{biblo}

\end{document}